\documentclass[a4paper,twoside,10pt]{letter}
\usepackage{graphicx,saj,multicol,subeqnarray,amssymb}

\def\papertype{ \ \hfill\ Preliminary
report}

\def\udc{524.37--77--54}
\begin{document}
\baselineskip=3.1truemm
\columnsep=.5truecm
\newenvironment{lefteqnarray}{\arraycolsep=0pt\begin{eqnarray}}
{\end{eqnarray}\protect\aftergroup\ignorespaces}
\newenvironment{lefteqnarray*}{\arraycolsep=0pt\begin{eqnarray*}}
{\end{eqnarray*}\protect\aftergroup\ignorespaces}
\newenvironment{leftsubeqnarray}{\arraycolsep=0pt\begin{subeqnarray}}
{\end{subeqnarray}\protect\aftergroup\ignorespaces}


\markboth{\eightrm THE $\Sigma-D$ RELATION FOR PLANETARY NEBULAE}
{\eightrm D. URO{\v S}EVI{\' C} {\eightit{\lowercase{et al.}}}}

{\ }

\publ

\type

{\ }

\title{THE $\Sigma-D$ RELATION FOR PLANETARY NEBULAE:\break
PRELIMINARY ANALYSIS}


\authors{D. Uro{\v s}evi{\' c}$^{\bf 1}$, B. Vukoti{\'c}$^{\bf 2}$, B.
Arbutina$^{\bf 1,2}$ and D. Ili{\'c}$^{\bf 1}$}

\vskip3mm

\address{$^1$Department of Astronomy, Faculty of Mathematics,
University of Belgrade\break Studentski trg 16, 11000 Belgrade,
Serbia}

\address{$^2$Astronomical Observatory, Volgina 7, 11160 Belgrade 74, Serbia}


\dates{February 22, 2007}{March 30, 2007}


\summary{An analysis of the relation between radio surface
brightness and diameter, so-called $\Sigma-D$ relation, for
planetary nebulae (PNe) is presented: i) the theoretical
$\Sigma-D$ relation for the evolution of bremsstrahlung surface
brightness is derived; ii) contrary to the results obtained
earlier for the Galactic supernova remnant (SNR) samples, our
results show that the updated sample of Galactic PNe
does not severely suffer from volume selection effect - Malmquist
bias (same as for the extragalactic SNR samples) and; iii) we
conclude that the empirical $\Sigma-D$ relation for PNe derived in
this paper is not useful for valid determination of distances for
all observed PNe with unknown distances.}



\begin{multicols}{2}
{


\section{1. INTRODUCTION}

\vskip2mm

The relation between radio surface brightnesses and diameters of
supernova remnants (SNRs), the so-called $\Sigma-D$ relation, has
been subject of the extensive discussions in the last more than
fourty years. Due to improvements of the observational techniques
(radio-interferometers), the several hundreds planetary nebulae
(PNe) were resolved in the last two decades at radio frequencies,
but the $\Sigma-D$ relation for PNe was not discussed until now.
By using radio data, some statistical methods were established in
order to determine distances to PNe. The main method was related
to the correlation between radius of PNe and brightness
temperature -- $R-T_{\rm b}$ relation (Van de Steene and Zijlstra
1995, Zhang 1995, Phillips 2002). The different samples of
Galactic PNe with known distances were defined in these papers.
All the obtained empirical $R-T_{\rm b}$ relations were used
for determination of distances to PNe for which the independent
distances (in order of $R-T_{\rm b}$ dependence) were not obtained
earlier.


The samples of Galactic PNe are better for statistical analysis
than the samples of Galactic SNRs. The selection effects should be
smaller in the case of PN samples. However, the selection effects
surely influence the Galactic PN samples and the statistical
determination of distances to Galactic PNe has to be highly
uncertain.

\smallskip

The main objectives of this paper are the following:

i) to derive a simple form of the theoretical $\Sigma-D$ relation
for PNe by analyzing the evolution of radio bremsstrahlung surface
brightness,

ii) to discuss whether the updated sample of radio PNe is
affected by the selection effects, and,

iii) to check whether the $\Sigma-D$ relation is valid for
determination of distances to PNe.
}

\end{multicols}

\vfill\eject

\begin{multicols}{2}

{
\section{2. ANALYSIS AND RESULTS}

\vspace{-.5cm}

\subsection{2.1. Theoretical $\Sigma-D$ relation for PNe}

The thermal bremsstrahlung mechanism is responsible for radiation
of HII regions at radio wavelengths. The bremsstrahlung volume
emissivity $\varepsilon_\nu$ of a PN can be shown to be (Rohlfs and
Wilson 1996):

\begin{equation}
\varepsilon_\nu [\mbox{ergs~s$^{-1}$~cm$^{-3}$~Hz$^{-1}$}] \propto
{n^2 T^{-1/2}},
\end{equation}

\noindent where $n$ is the volume density and $T$ is the
thermodynamic temperature of interstellar medium (ISM).

The surface brightness can be expressed as:

\begin{equation}
\Sigma_\nu\propto\varepsilon_\nu D,
\end{equation}

\noindent where $D$ is the diameter of PN. Combining Eqs. (1) and
(2), we obtain:

\begin{equation}
\Sigma_\nu\propto n^2T^{-1/2} D.
\end{equation}

\noindent Our next step is to express dependance of $n$ and $T$ on
$D$. For a constant velocity mass flow the density distribution is
$\varrho={\dot{\mathcal{M}}\over 4\pi r^2v}$, i.e. $n\propto
D^{-x}$, where $x=2$. Moreover, for the isothermal envelope
with a power-law electron density distribution there is
relationship between the shape of the density distribution and the
power-law index of the radio continuum spectra (see Gruenwald and
Aleman 2007, and references therein). Supposing that $n\propto
D^{-2}$ and $T$=const. (HII regions are approximately isothermal
at $T\sim10^4$ K), we obtain the simplest form of the theoretical
$\Sigma-D$ relation for PNe:

\begin{equation}
\Sigma_\nu\propto D^{-3}.
\end{equation}

\noindent This is a standard power-law form of the $\Sigma-D$
relation which can be written in general form as
$\Sigma=AD^{-\beta}$, that is the same as in the case of SNRs.

It is possible that $x$ in density distribution is slightly
higher, $x\gtrsim 2$, and that the temperature is not strictly
constant throughout the nebula. We can expect to see temperature
gradients in PNe arising from radiation hardening. More energetic
photons will travel further and when they are absorbed by the PN
they will impart greater kinetic energy to the ions thereby
producing a higher temperature. Using the numerical model results
given by Evans and Dopita (1985), we calculate the dependence
between $\log T_{\rm e}$ and $\log D$ and find the low slope
($\approx 0.1$). Therefore, this only slightly changes the slope
of the theoretical $\Sigma-D$ relation. The value $\beta = 3$ is
then a theoretical lower limit, and the $\Sigma-D$ relation could
only be steeper, as one can see from Eq. (3).

\subsection{2.2. The empirical $\Sigma-D$ relation for PNe}

The most important prerequisite for deriving a proper empirical
$\Sigma-D$ relation is defining of a representative sample of PNe. The
distances to the calibrators have to be determined by accurate
methods, e.g. trigonometric or spectroscopic parallaxes of
central stars in PNe, or by a method that uses the expansion of
nebulae. On the other hand, all samples suffer from the severe
selection effects that arise from limitation in sensitivity and
resolution, but the most severe selection effect for the Galactic
samples of PNe is Malmquist bias; i.e. intrinsically bright PNe
are favored because they are sampled from a larger spatial volume
compared to any given flux limited survey. The result is a bias
against low surface brightness nebulae such as highly evolved old
PNe. In this paper we use the updated sample of PNe at the
distances less than 0.7 kpc collected by Phillips (2002). The
influence of Malmquist bias in this sample is limited because of
the limitation in distances to PNe. In addition, we assume that the
distances are accurately determined for this sample of relatively
close PNe. The empirical $\Sigma-D$ relation at 5 GHz for 44
calibrators with distances less than 0.7 kpc (Phillips 2002) has
the form:

\begin{equation}
\Sigma_\mathrm{56 Hz}=2.33^{+0.88}_{-0.64}\cdot10^{-22}D^{-2.07\pm0.19}.
\end{equation}

\noindent The parameters $A$ and $\beta$ are calculated by
least-squares fitting procedure with correlation coefficient
$-0.86$. The corresponding $\Sigma_\nu-D$ diagram is shown in Fig. 1.

\vskip 9mm

\centerline{\includegraphics[width=8cm]{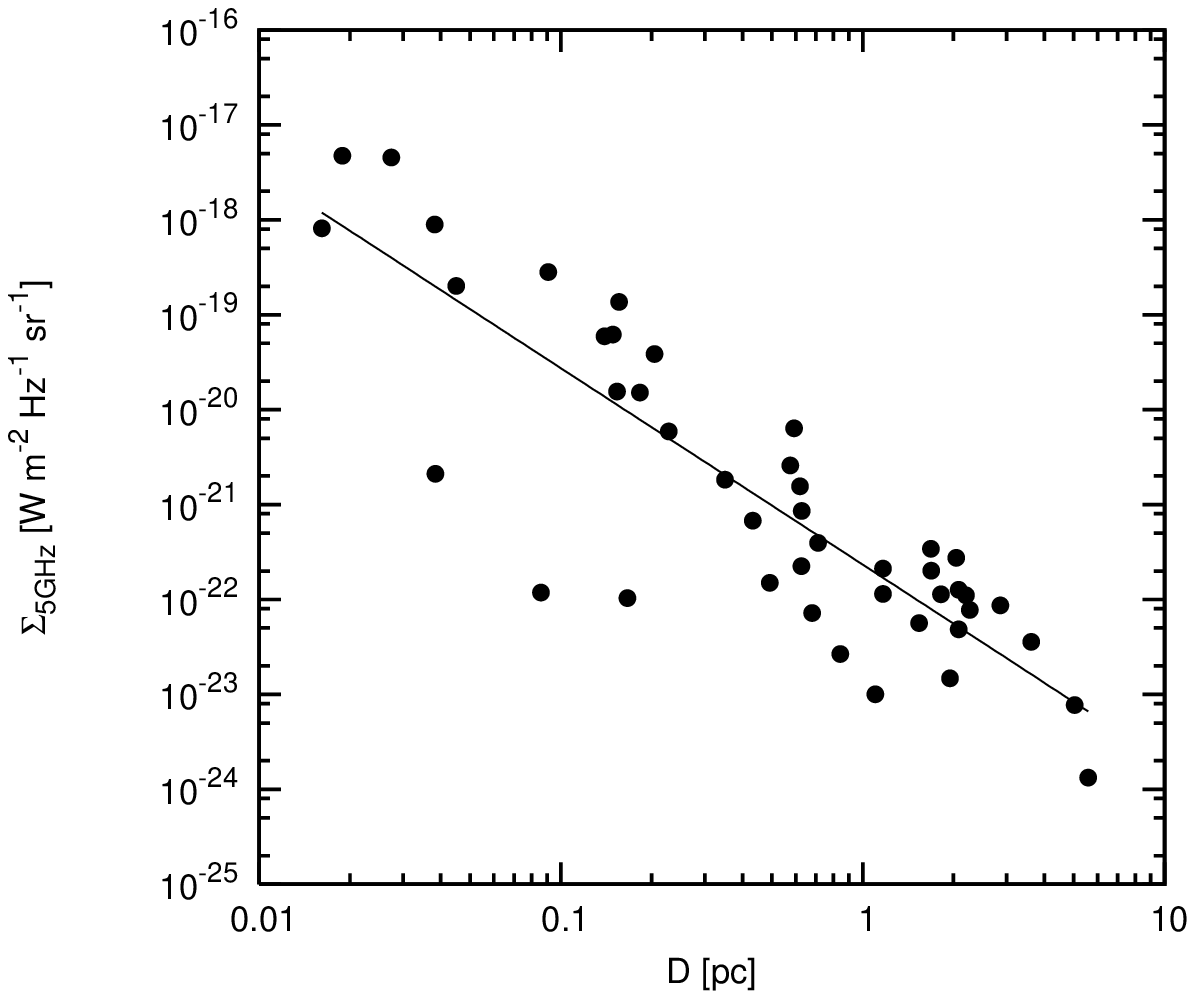}}


\figurecaption{1.}{The $\Sigma-D$ diagram at 5 GHz for 44 Galactic
PNe with distances less than 0.7 kpc.}


The form of Eq. (5) is very close to the so-called trivial $\Sigma-D$
form with $\beta=2$ (for details see Arbutina et al. 2004). The
additional test in order to estimate the validity of Eq. (5)
pertains to the possible dependence between the luminosity and
diameter of PNe. The $L_\nu-D$ diagram is shown in Fig. 2. The scatter
in $L_\nu-D$ plane shows that the correlation between $L_\nu$ and $D$ is
poor (correlation coefficient = -0.06) and therefore the physical
dependence between $L$ and $D$ could not be confirmed by this
statistical procedure.

\vskip 6mm

\centerline{\includegraphics[width=8cm]{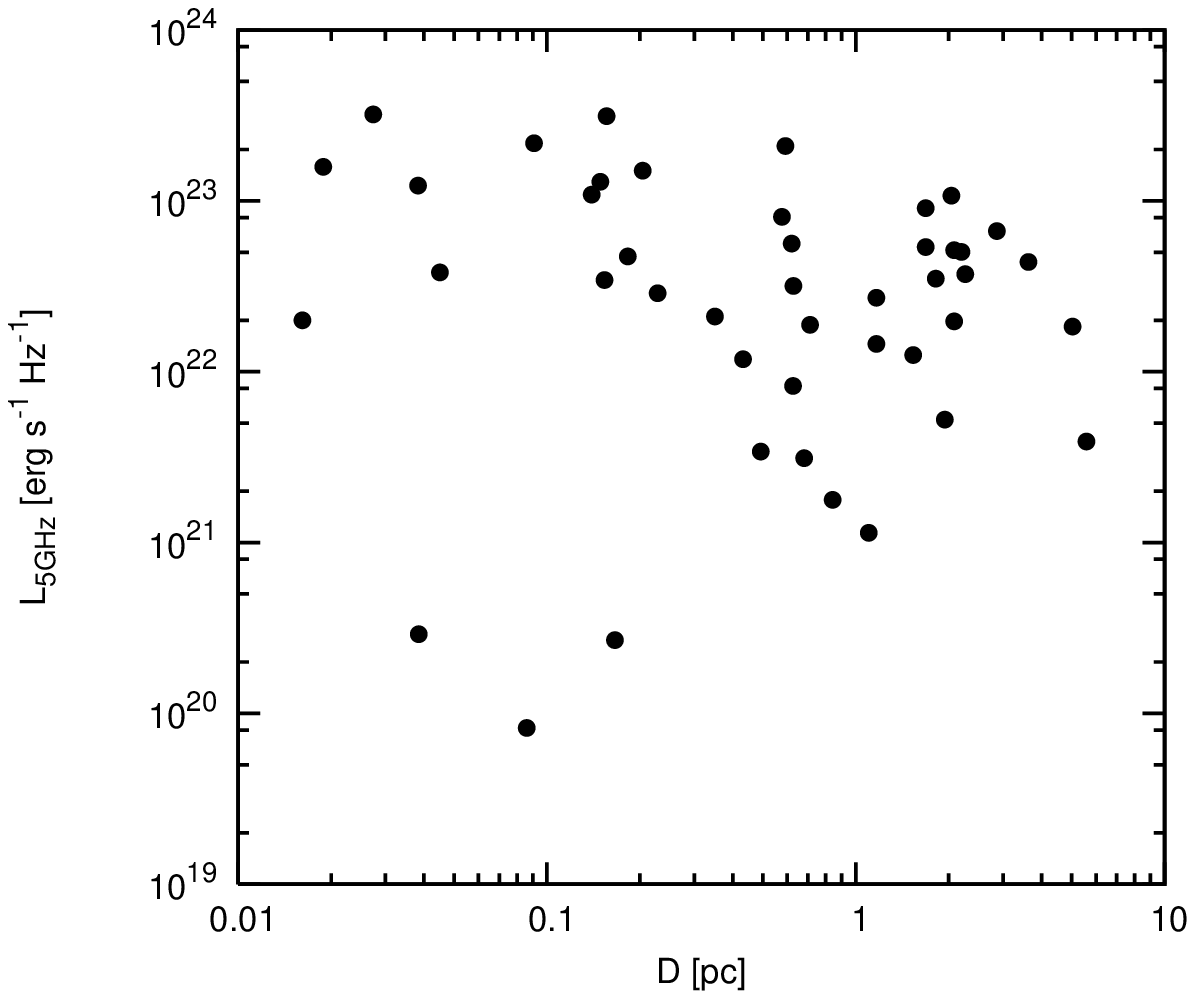}}


\figurecaption{2.}{The $L-D$ plot at 5 GHz for 44 Galactic PNe
with distances less than 0.7 kpc.}


\section{3. DISCUSSION}

The theoretical $\Sigma_\nu-D$ relation (Eq. (4)) for PNe, derived in
this paper, describes a trend of decreasing radio surface
brightness with increasing diameter of an object. The radiation
mechanism used in this simple derivation is thermal
bremsstrahlung. This is the basic process of production of the
radio radiation in HII regions. The theoretically derived slope
($\beta=3$) is steeper than the slope from the empirical relation
given by Eq. (5). This discrepancy can be explained by the low
quality of the sample of Galactic PNe or by the assumptions used
in derivation of theoretical relation. Due to small variation in
power-law density distribution with $x\gtrsim 2$ (Gruenwald and
Aleman 2007, and references therein) and approximately constant
temperature of expanding envelope of PNe, theoretical slope can be
slightly steeper than in Eq. (4). Therefore, we conclude that the
theoretical relation has the correct form, but our empirical
relation is under influence of biases that could make the slope
shallower. On the other hand, there are some attempts to show
that evolution of PNe are not linear in log-log scales (e.g.
Phillips 2004). These different dependences cannot be derived
from the thermal bremsstrahlung radiation formula (Eq. (1)).

A very interesting feature regarding the empirical relation
for Galactic PNe (Eq. (5)) is that the slope is approximately
equal to the slope of trivial $\Sigma-D$ relation. Therefore, we
conclude that Malmquist bias is not so severe as in cases of
Galactic SNR samples. This slope ($\beta\approx2$) was obtained
for the extragalactic samples of SNRs (except M82 sample) where
Malmquist bias is small, because all the SNRs are
at the approximately same distance (see Uro{\v s}evi{\' c} 2002,
Uro{\v s}evi{\' c} et al. 2005).

The large scatter in $L_\nu-D$ plane (Fig. 2)
suggests that the slope in Eq. (5) does not have real and valid
physical interpretation. It is a kind of luminosity-diameter
scattering artefact which produces the trivial $\Sigma\propto
D^{-2}$ form. Therefore, the relation defined by Eq. (5) is not
precise enough for determination of valid distances to Galactic
PNe. This is due to the different biases: the limitations in
sensitivity and resolution of radio surveys, the source confusion,
Malmquist bias (in mild form), mixture of different types of PNe
in the same sample, and insufficient precision in determining the distances
to the 44 calibrators.

\section{4. SUMMARY}

The main results of this paper may be summarized as follows:

\bigskip

\item{i)} The theoretical $\Sigma_\nu-D$ relation for the radio
evolution of thermal bremsstrahlung surface brightness of PNe in
form of $\Sigma_\nu\propto D^{-3}$ is derived.

\smallskip

\item{ii)} Our results show that the updated sample of
Galactic PNe does not severely suffer from volume selection effect
- Malmquist bias (same as in cases of the extragalactic SNR
samples). This is opposite to results obtained
earlier for the Galactic SNR samples.

\smallskip

\item{iii)} Due to analysis of the $L_\nu-D$ dependence, we conclude
that the $\Sigma_\nu-D$ relation for Galactic PNe is not useful for
reliable determination of distances for all observed PNe with unknown
distances.

\bigskip

The above observation leads to the more general comment that PNe
may have very different initial conditions leading to
independent evolutionary paths. These paths could follow the same
theoretical $\Sigma-D$ curve but with varying intercepts,
leading to the scatter such as the one found in this paper.


\acknowledgements{The authors would like to thank the referee
Prof. Nebojsa Duric for valuable comments which have improved this
paper. This research has been supported by the Ministry of Science
and Environmental Protection of the Republic of Serbia (Projects:
No 146002, No 146003, No 146012, No 146016).}
}

\end{multicols}

\vfill\eject

\begin{multicols}{2}

{
\references

Arbutina, B., Uro{\v s}evi{\' c}, D., Stankovi{\' c}, M. and Te{\v
s}i{\' c}, Lj.: 2004, \journal{Mon. Not. R. Astron. Soc.},
\vol{350}, 346.

Evans, I.N. and Dopita, M.A.: 1985, \journal{Astrophys. J. Suppl.
Series}, \vol{58}, 125

Gruenwald, R. and Aleman, A.: 2007, \journal{Astron. Astrophys.},
\vol{461}, 1019.

Phillips, J.P.: 2002, \journal{Astrophys. J. Suppl. Series},
\vol{139}, 199.

Phillips, J.P.: 2004, \journal{Mon. Not. R. Astron. Soc.},
\vol{353}, 589.

Rohlfs, K. and Wilson, T.L.: 1996, Tools of Radio Astronomy,
Springer

Uro{\v s}evi{\' c}, D.: 2002, \journal{Serb. Astron. J.},
\vol{165}, 27

Uro{\v s}evi{\' c}, D., Pannuti, T. G., Duric, N.,  Theodorou, A.:
2005, \journal{Astron. Astrophys.}, \vol{435}, 437.

Van de Steene, G.C. and Zijlstra, A.A.: 1995, \journal{Astron.
Astrophys.}, \vol{293}, 541.

Zhang, C.Y.: 1995, \journal{Astrophys. J. Suppl. Series},
\vol{98}, 659.

\endreferences

}
\end{multicols}

{\ }



\naslov{$\Sigma-D$ RELACIJA ZA PLANETARNE MAGLINE: PRELIMINARNA ANALIZA}


\authors{D. Uro{\v s}evi{\' c}$^{\bf 1}$, B. Vukoti{\'c}$^{\bf 2}$, B.
Arbutina$^{\bf 1,2}$ and D. Ili{\'c}$^{\bf 1}$}

\vskip3mm


\address{$^1$Department of Astronomy, Faculty of Mathematics,
University of Belgrade\break Studentski trg 16, 11000 Belgrade,
Serbia}

\address{$^2$Astronomical Observatory, Volgina 7, 11160 Belgrade 74, Serbia}

\vskip.7cm


\centerline{\rrm UDK \udc}

\vskip1mm


\centerline{\rit Prethodno saopxte{\nj}e}

\vskip.7cm

\begin{multicols}{2}
{


\rrm Prikazana je analiza tzv.~$\Sigma-D$ re\-la\-ci\-je
izme{\dd}u povrxinskog s{\jj}a{\jj}a na radio-frekvenci{\jj}ama i
dijametra planetarnih maglina (PM): {\rm i)} izvedena {\jj}e
teorijska $\Sigma-D$ relacija za evoluciju povrxinskog sjaja
stvorenog zakoqnim zraqenjem; {\rm ii)} suprotno rezultatima
dobi{\jj}enim rani{\jj}e za uzorke saqi{\nj}ene od Galaktiqkih
ostataka supernovih, naxi rezultati pokazu{\jj}u da
na{\jj}\-no\-vi\-{\jj}e formirani uzorak Galaktiqkih PM ne trpi veliki
uticaj zbog zapreminskog selekcionog efekta, tzv.
Malmkvistovog selekcionog efekta (isto va{\zz}i za vangalaktiqke
uzorake ostataka supernovih); i {\rm iii)} zak{\lj}uqu{\jj}emo da
empirijska $\Sigma-D$ relacija za PM izvedena u ovom radu
nije upotreb{\lj}iva za pouzdana odre{\dd}iva{\nj}a da{\lj}ina do
svih posmatranih PM sa nepoznatim daljinama.
}

\end{multicols}

\end{document}